# The impact of class imbalance in logistic regression models for low-default portfolios in credit risk

Willem D Schutte[1,2] , Charl Pretorius[1,2] , Neill Smit[1,2] , Leandra van der Merwe[1] , Robert Maxwell[1]

[1]*Centre for Business Mathematics and Informatics & Unit for Data Science and Computing, North-West University, South Africa*

[2]*National Institute for Theoretical and Computational Physics (NITheCS), Potchefstroom, South Africa*

## Abstract

In this paper, we study how class imbalance, typical of low-default credit portfolios, affects the performance of logistic regression models. Using a simulation study with controlled data-generating mechanisms, we vary (i) the level of class imbalance and (ii) the strength of association between the predictors and the response. The results show that, for a given strength of association, achievable classification accuracy deteriorates markedly as the event rate decreases, and the optimal classification cut-off shifts with the level of imbalance. In contrast, the Gini coefficient is comparatively stable with respect to class imbalance once sample sizes are sufficiently large, even when classification accuracy is strongly affected. As a practical guideline, we summarise attainable classification performance as a function of the event rate and strength of association between the predictors and the response.

Keywords: Class imbalance; Classification; Concordance; Credit risk; Credit scorecards; Gini coefficient; Logistic regression; Weight of evidence

## 1   Introduction

The estimation of the risk parameters for portfolios that exhibit a low-default nature poses a common challenge for financial institutions worldwide. For these types of portfolios, typically referred to as low-default portfolios (LDPs), default events are very limited and extremely underrepresented in the data, causing the event and nonevent classes to be imbalanced. Such data sets are said to exhibit *class imbalance*. When banks develop models for risk parameters based on LDPs, it is not known precisely what impact the class imbalance problem will have on aspects of the model performance, such as the accuracy of predictions, the discrimination ability of the model, and classification cut-offs. It is also known that regulators prefer models to be developed on real client data rather than simulated data.

When using logistic regression in classification problems, it is well documented that class imbalance can have detrimental effects, including a low classification accuracy for the minority class, weak separation ability of the classifier, and underestimation of the probability of a rare event. Standard classifiers are often biased towards the majority class and consequently have a higher misclassification rate for the minority class (López et al., 2013). Furthermore, the choice of performance metric guiding the learning process might induce further bias if the performance metric does not adequately penalise misclassification of the minority class (Haixiang et al., 2017). Zhang et al. (2018) show that, when trained on heavily imbalanced data sets, logistic regression can yield a very small false negative rate while having a very high false positive rate. A numerical study by Zhang et al. (2018) indicates that the separation ability, as measured by the Kolmogorov-Smirnov statistic, of the logistic regression classifier can be sensitive to a low event rate. This could be caused by minority class observations overlapping with majority class observations (Haixiang et al., 2017). According to King and Zeng (2001), applying logistic regression to small samples can lead to underestimation of the probability of the rare event. In the case of a more severe class imbalance in the data set, a greater bias in the predicted probabilities can be expected (Puhr et al., 2017). It should be noted that studies that investigate the effects of class imbalance are typically based on real data sets. There is a clear lack of controlled simulation studies to properly evaluate the impact of imbalanced data, especially in the credit risk environment.

There exist several remedial measures to address adverse effects introduced by the class imbalance problem. It is apparent from the surveyed literature that it is still unclear which combination of classification model and remedial measure is most appropriate in a given situation (see, e.g., Jiang et al., 2023, van den Goorbergh et al., 2022, Xiao et al., 2021, López et al., 2013). We briefly describe four categories of remedial measures: resampling approaches, synthesising techniques, embedded framework approaches, and post-hoc adjustment methods. This taxonomy is proposed by Jiang et al. (2023), which is a refinement of what is usually referred to as data-level techniques and algorithm-level techniques (see, e.g., Rout et al., 2018).

Resampling approaches involve resampling the data to reduce class imbalance at a data level. Popular techniques include random over-sampling, random under-sampling, and adapted versions of these two techniques (see Jiang et al., 2023). Resampling approaches appear to be the most popular in credit scoring applications, most likely due to the simplicity and ease of implementation. However, there are contradicting results on the efficacy of resampling approaches in the literature, and most studies are based on real data sets instead of controlled simulations settings. Kim and Hwang (2022) show that resampling techniques can be ineffective in improving a classifier's performance and can in some cases also be detrimental. van den Goorbergh et al. (2022) show that, after correcting for class imbalance, the probability of belonging to the minority class is significantly overestimated. Marqués et al. (2013) demonstrate that using resampling methods leads to a consistent improvement in model performance, with over-sampling performing better than any under-sampling approach. A relabelling approach was recently proposed by Li (2020), where the author shows that resampling the minority class is insufficient for handling the class imbalance problem and demonstrates that the proposed relabelling approach is more effective in capturing the underlying structure of the minority class. Synthesising techniques are used to generate additional minority class observations using mechanisms that avoid the repetition or duplication of entries from the minority class. Popular synthesising techniques include synthetic minority over-sampling technique (SMOTE; see Chawla et al., 2002) and its various extensions, generative adversarial networks (GANs; see Goodfellow et al., 2014), and the adaptive synthetic sampling method (ADASYN; see He et al., 2008).

Embedded framework approaches refer to techniques that involve integrating resampling methods or synthesising techniques into existing model-building approaches not inherently suitable for class-imbalanced data. Two examples are the sampling-embedded approach, which involves combining, for example, under-sampling with bagging or SMOTE with boosting, and the cost-embedded approach, which combines resampling approaches with cost-sensitive learning where misclassifications are penalised according to the severity of the consequence of such misclassifications.

Jiang et al. (2023) mention two post-hoc adjusting methods: threshold adjusting and MetaCost. Threshold adjusting involves modifying the classification cut-off to minimise the total cost of misclassifying observations. This approach is appropriate when the cost of misclassification is asymmetrical, e.g., when the cost of approving a loan to a customer who is likely to default is greater than the cost of not approving a loan to a customer who is likely to not default. For a description of MetaCost, which is related to cost-sensitive classification, see Domingos (1999).

Besides the event rate, which López et al. (2013) indicate as the main factor affecting classifier performance, most studies find that other prominent factors such as sample size, number of predictors, and number of events per predictor variable also influence model performance. The reader is referred to López et al. (2013) for further discussions on classification problems in the presence of a low event rate.

It is expected that the number of observations available for training a model will affect its performance. In contrast to the relatively small sample sizes typically recommended for credit scoring applications (see, e.g., Siddiqi, 2017), Crone and Finlay (2012) advocate the use of samples containing at least 1 500 to 2 000 observations. Traditional maximum likelihood estimation is known to have potentially problematic small-sample behaviour (van Smeden et al., 2019, lists five specific issues). As a remedy, van Smeden et al. (2019) shows that regression shrinkage can alleviate these adverse effects. They also suggest that the required sample size for a given situation should be determined according to out-of-sample predictive performance.

Based on a simulation study involving resampling from clinical trial data, Peduzzi et al. (1996) showed that for situations where the number of minority-class events per predictor variable, or simply referred to as events per variable (EPV), was less than 10, the regression coefficients were biased in both directions and the sample variances of the estimated coefficients were overestimated. In fact, several other authors suggest an EPV of at least 10 (see, e.g., Harrell Jr et al., 1985, Wynants et al., 2015), and Freedman and Pee (1989) showed that an EPV of less than 4 leads to increased overfitting. In contrast to these findings, van Smeden et al. (2019) found that EPV does not have a strong relation with metrics of predictive performance. They claim that EPV is not a suitable criterion for binary prediction model development. Instead, they recommend considering the number of predictors, the total sample size, and the event rate.

It is expected that certain predictors will contain more predictive value about the event than other predictors. In addition, even if several predictors are combined in a model, there will still be an unobserved error component causing unexplained variability in the true outcome. If the error variability is so large that it obfuscates the information contained in the predictors, the noise (error) is more prominent than the signal (predictors), and we say that there is a low signal-to-noise ratio. If the error is negligible and the predictors contain clear information regarding the outcome, we have a high signal-to-noise ratio. To our knowledge, there exists no literature on the effect of the signal-to-noise ratio in the case of class imbalance. Most literature focuses on proxies such as EPV mentioned above.

In this paper, we consider the impact of class imbalance in logistic regression models for credit risk in a controlled simulation study setup. Several simulation configurations are considered under different levels of class imbalance. In line with the findings of van Smeden et al. (2019), we consider as criteria for predictive performance the sample size and event rate. However, we use an aggregate information value (AIV), which takes into account the number of predictors and predictive value of each predictor, to capture the signal-to-noise ratio. No remedial measures are applied, since the goal of this study is to evaluate the impact of class imbalance in this context. We will also show, when using imbalanced training data sets, training samples should be large enough, as advocated by Crone and Finlay (2012).

The remainder of the paper is structured as follows. In Section 2, we discuss several methodological considerations, including preliminaries for the simulation study and performance metrics. Section 3 describes the simulation setup tailored to be representative of typical imbalanced data found in credit risk modelling, with the results of the simulation study provided and discussed in Section 4. In Section 5, we present practical guidelines, closing remarks, and recommendations for future research.

## 2 Methodological framework

### 2.1 Modelling preliminaries

In credit scoring, it is common practice to discretise continuous predictor variables into discrete bins. Suppose that we have $d$ discrete (or "binned") predictors, denoted by $X_j$, $j = 1, \ldots, d$, which may be related to the outcome of an event. The event indicator variable $Y$ has mass function

$$P(Y = k) = \pi_1^k (1 - \pi_1)^{1-k}, \qquad k \in \{0, 1\},$$

where $\pi_1 \in (0, 1)$ denotes the *event rate*. That is, $\pi_1$ denotes the probability that the event $Y = 1$ occurs. Also define the nonevent rate $\pi_0 = 1 - \pi_1$.

A common approach in credit risk modelling involves using weights of evidence (see, e.g., Siddiqi, 2017). The weight of evidence of a predictor measures the strength of a predictor in distinguishing between events and nonevents. The weight of evidence associated with the event $X_j = x$ is defined by

$$w_j(x) = \ln \frac{P(X_j = x | Y = 0)}{P(X_j = x | Y = 1)}. \tag{2.1}$$

The quantity in (2.1) is unknown and is typically estimated using the adjusted estimator

$$\widehat{w}_j(x) = \ln \frac{[\#(X_j = x, Y = 0) + \vartheta] / \#(Y = 0)}{[\#(X_j = x, Y = 1) + \vartheta] / \#(Y = 1)},$$

where $\#(A)$ denotes the number of times the event $A$ occurred in the sample, and $\vartheta$ is an arbitrary adjustment factor incorporated to avoid the estimator being undefined when $\#(X_j = x, Y = 1) = 0$. Throughout, we choose $\vartheta = 0.5$ because this is the default choice in SAS and therefore used by many practitioners (see SAS Institute Inc., 2018, p. 1097).

The idea underlying the use of weight of evidence is grounded in the relationship between conditional probabilities and odds, which will now be presented briefly. For a more detailed discussion on the use of weights of evidence in the context of credit scorecards, the reader can consult Thomas (2009).

Typically, one would model the relationship between $Y$ and $\boldsymbol{X}$ using a logistic regression model where the response variable is the conditional log odds of the event $Y = 1$ given $\boldsymbol{X} = \boldsymbol{x}$, that is,

$$\operatorname{logit} P(Y = 1 | \boldsymbol{X} = \boldsymbol{x}) = \ln \frac{P(Y = 1 | \boldsymbol{X} = \boldsymbol{x})}{P(Y = 0 | \boldsymbol{X} = \boldsymbol{x})}.$$

Invoking Bayes' rule, and assuming that the $d$ components of $\boldsymbol{X} = (X_1, \ldots, X_d)$ are conditionally independent (conditional on $Y$), the conditional log odds can be written in terms of the weights of evidence as

$$\begin{aligned}
\operatorname{logit} P(Y=1|X=x) &= \log\frac{P(Y=1)}{P(Y=0)} + \log\frac{P(X_1=x_1|Y=1)\cdots P(X_d=x_d|Y=1)}{P(X_1=x_1|Y=0)\cdots P(X_d=x_d|Y=0)} \\
&= \operatorname{logit} P(Y=1) - \sum_{j=1}^{d} w_j(x_j).
\end{aligned} \quad (2.2)$$

This shows that, given attributes $X = x$, the conditional log odds of observing $Y = 1$ is perfectly negatively correlated with the weight of evidence corresponding to the event $X = x$. This means that, given $X = x$, greater values of the weight of evidence are associated with lower chances of observing the event $Y = 1$.

In practice, it is possible that the predictors are not independent. As an alternative approach, $\operatorname{logit} P(Y = 1|X = x)$ could be modelled using the more general logistic regression model

$$\pi(x; \boldsymbol{\beta}) = \beta_0 + \sum_{j=1}^{d} \beta_j w_j(x_j), \quad (2.3)$$

where $\boldsymbol{\beta} = (\beta_0, \dots, \beta_d)$ is an unknown parameter vector to be estimated from data.

## 2.2  Information value

For each $j = 1, \dots, d$, assume that the regressor $X_j$ can take on one of $K_j$ possible values ("bins") $a_{j1}, \dots, a_{jK_j}$. The *information value* (IV) associated with $X_j$ is defined as

$$\mathcal{I}_Y(X_j) = \sum_{k=1}^{K_j} \left[P(X_j = a_{jk}|Y=0) - P(X_j = a_{jk}|Y=1)\right] w_j(a_{jk}), \quad j = 1, \dots, d. \quad (2.4)$$

Higher values of $\mathcal{I}_Y(X_j)$ means that the attribute $X_j$ carries information on the distribution of the event $Y$. Note that if $X_j$ and $Y$ are independent, then $\mathcal{I}_Y(X_j) = 0$. The information value originates from information theory and is related to relative Shannon entropy (Shannon, 1948), also known as the Kullback–Leibler divergence (Kullback and Leibler, 1951).

For this study, it will also be of interest to determine the distributional information on $Y$ that can be determined from knowledge of a vector of predictors $X = (X_1, \dots, X_d)$, and vice versa. To this end, we define the *aggregate information value* (AIV) associated with $X$ by

$$\mathcal{I}_Y(X) = \sum_{a \in \mathcal{D}} \left[ \left(p_{X|Y}(a|0) - p_{X|Y}(a|1)\right) \ln \frac{p_{X|Y}(a|0)}{p_{X|Y}(a|1)} \right], \quad (2.5)$$

where $\mathcal{D}$ denotes the set of possible values of $X$. Throughout the paper, the AIV represents the degree/strength of association between the response variable $Y$ and a set of predictors $X = (X_1, \dots, X_d)$. Under conditional independence of the components of $X$ given $Y$, it follows that $\mathcal{I}_Y(X) = \sum_{j=1}^{d} \mathcal{I}_Y(X_j)$, with $\mathcal{I}_Y(X_j)$ as defined in (2.4). That is, if the predictors are conditionally independent of one another, the AIV is simply the sum of the individual IVs.

**Remark.** The information values defined in (2.4) and (2.5) are related to the population stability index frequently used in credit scoring in the sense that they are also symmetrised versions of the Kullback-Leibler distance; see, e.g., Lewis (1994), Potgieter et al. (2025), Siddiqi (2017). However, while the population stability index is typically used to compare the development population with the testing population, our focus here is on comparing, for each predictor, its conditional distribution within the event class to its conditional distribution within the nonevent class.

## 2.3 Performance metrics

### 2.3.1 Classification metrics

Classification metrics, also referred to as threshold-dependent metrics, are a class of performance metrics derived from the confusion matrix; Table 1 shows a confusion matrix for binary classification problems. After reviewing existing literature on classification metrics and their use in credit risk modelling, we consider only two classification metrics in this paper.

*Table 1. Binary confusion matrix.*

|  |  | Predicted | |
|---|---|---|---|
|  |  | Positive | Negative |
| Actual | Positive | True positive (TP) | False negative (FN) |
|  | Negative | False positive (FP) | True negative (TN) |

The $F_1$ score is one of the most widely used classification metrics and is given by the harmonic mean of recall and precision, which can be simplified to

$$F_1 \text{ score} = \frac{2\text{TP}}{2\text{TP} + \text{FP} + \text{FN}}.$$

The $F_1$ score has been extensively used in studies with class imbalances, but recent research has indicated that the metric might not always be appropriate in certain cases of class imbalance. This is of course evident from the definition of the metric, where equal weights are assigned to recall and precision. Recall represents the proportion of correctly predicted positives out of the actual positives, where precision represents the proportion of correctly predicted positives out of all predicted positives. The $F_1$ score thus focuses only on correctly classifying the positive class, which is typically the minority class representing defaults in the credit scoring environment. Since the cost of an incorrectly classified default is typically much higher than the cost of an incorrectly classified non-default, the $F_1$ score should still be a suitable metric for measuring classifier performance on imbalanced credit data.

The $P_4$ score, recently proposed by Sitarz (2023), is a more comprehensive metric for handling imbalanced data. The $P_4$ score is a symmetrical extension of the $F_1$ score and is given by the harmonic mean of the recall, precision, specificity, and negative predictive value, which simplifies to

$$P_4 \text{ score} = \frac{4\text{TP} \times \text{TN}}{4\text{TP} \times \text{TN} + (\text{TP} + \text{TN})(\text{FP} + \text{FN})}.$$

Considering the definition of this metric, equal weight is given to the cost of misclassifying both the positive and negative classes. The $P_4$ score allows for label switching and its level of interpretability is similar to that of the $F_1$ score. Sitarz (2023) states that the $P_4$ score penalises more strongly than other comprehensive classification metrics such as the Matthews correlation coefficient (MCC). Furthermore, a weighted version of the $P_4$ score could also be considered (similar to a weighted version of the $F_1$ score) to give more weight to the class of interest. This is related to cost-based approaches, which can more heavily penalise misclassification of the class of interest.

In credit risk modelling, the positive class (the event class corresponding to defaults) is always the minority class and class of interest, thus both the $F_1$ score and $P_4$ score could be appropriate as classification metrics.

### 2.3.2  Discrimination ability

The Gini coefficient is a popular measure used in credit scoring to assess the degree of concordance between predictions of the probability of default obtained from a classifier and the actual occurrence of the default event. That is, given two customers, a classifier with a high Gini score will in most cases be able to correctly identify which of the two customers are more likely to default. The Gini coefficient is therefore useful when ranking customers is more important than predicting whether the customer will actually default or not.

The Gini coefficient, also known as Somers' $D$ (see, for example, Newson, 2002), can be used to measure the degree of concordance between $n$ predictions $\hat{Y}_1, \ldots, \hat{Y}_n$ obtained from a fitted logistic regression model and the corresponding true responses $Y_1, \ldots, Y_n$. The Gini coefficient of $\hat{Y}$ with respect to $Y$ is defined by $D_{\hat{Y}|Y} = (n_c - n_d)/n_u$, where $n_c$ is the number of concordant pairs, $n_d$ is the number of discordant pairs, and $n_u$ is the number of pairs with different responses (i.e., not tied in the response). If all pairs in the sample are concordant, then the Gini coefficient is equal to 1. If all pairs are discordant, then the Gini is equal to –1.

We finally mention that the Gini coefficient is closely related to the rank-based correlation measure known as Kendall's tau (Kendall, 1938). For more details on this relation, the interested reader is referred to Newson (2002).

## 3   Simulation setup

### 3.1   Simulation procedure

We conduct a Monte Carlo simulation consisting of 500 Monte Carlo iterations for each configuration. The sample sizes considered are 50, 100, ..., 450, 500, 750, 1 000, 1 500, 2 000 and 2 500. Fixed event rates of 1%, 5%, and 10% are considered to investigate the effect of class imbalance on classification accuracy.

We now outline the procedure to generate a sample of size $n$ with a specified number of event cases, say $n_1 = \lfloor \pi_1 n \rfloor$, so that the event rate is $\pi_1$. To do this, set the first $n_1$ responses $Y_1, \ldots, Y_{n_1}$ equal to 1 (event cases) and the remaining $n - n_1$ responses $Y_{n_1+1}, \ldots, Y_n$ equal to 0 (nonevent cases). For each response $Y_i$, $i = 1, \ldots, n$, generate a $d$-dimensional vector of predictors $X_i = (X_{i,1}, \ldots, X_{i,d})^\top$ from the class-conditional probability distribution $P(X = x|Y = Y_i)$. The sample, which we denote by $(\mathcal{X}, Y)$, then consists of the design matrix $\mathcal{X} = (X_1, \ldots, X_n)^\top$ and the response vector $Y = (Y_1, \ldots, Y_n)^\top$.

Seeing that the classifier is a logistic regression model which predicts an event probability, and we are interested in classifying observations as an event or a nonevent, a classification cut-off point $\theta \in (0,1)$ needs to be chosen. This cut-off point is chosen by maximising the classification accuracy according to some classification metric, which we choose to be either the $F_1$ score or the $P_4$ score as defined in Section 2.3.1. Given a cut-off point of $\theta$, we denote by $C_\theta(\widehat{Y}|Y)$ the value of the chosen classification metric calculated using the predicted probabilities $\widehat{Y}$ and the actual binary observations $Y$.

The steps followed in each Monte Carlo iteration are listed below. The steps are also outlined visually in Figure 1.

1. Generate a training sample $(\mathcal{X}^{\text{train}}, Y^{\text{train}})$, a validation sample $(\mathcal{X}^{\text{val}}, Y^{\text{val}})$, and a test sample $(\mathcal{X}^{\text{test}}, Y^{\text{test}})$ according to the sampling procedure above. Note that the three samples each contain $n$ observations of which $n_1$ are events.
2. Calculate the weights of evidence from the training sample $(\mathcal{X}^{\text{train}}, Y^{\text{train}})$.
3. Using the training data, fit a logistic regression model by determining the parameter vector that maximises the conditional log-likelihood of $Y$ given $X$, i.e.,

$$\widehat{\boldsymbol{\beta}} = \underset{\boldsymbol{\beta} \in \mathbb{R}^d}{\operatorname{argmax}} \sum_{i=1}^{n} \left[ Y_i^{\text{train}} \ln \pi(X_i^{\text{train}}; \boldsymbol{\beta}) + \left(1 - Y_i^{\text{train}}\right) \ln(\pi(X_i^{\text{train}}; \boldsymbol{\beta})) \right],$$

where $\pi(x; \boldsymbol{\beta}) = \beta_0 + \sum_{j=1}^{d} \beta_j w_j(x_j)$ is as defined in (2.3).

4. Calculate the predictions $\widehat{Y}_i^{\text{val}} = \pi(X_i^{\text{val}}; \widehat{\boldsymbol{\beta}})$, $i = 1, \ldots, n$, by applying the estimated logistic regression model to the validation data.
5. Using the validation data and the predictions obtained in Step 4, determine the optimal cut-off $\widehat{\theta} \in (0, 1)$ that maximises the classification metric $C_\theta(\widehat{Y}^{\text{val}}|Y^{\text{val}})$.
6. Calculate the test set predictions $\widehat{Y}_i^{\text{test}} = \pi(X_i^{\text{test}}; \widehat{\boldsymbol{\beta}})$, $i = 1, \ldots, n$, by applying the estimated logistic regression model to the test set.
7. Determine $C_{\widehat{\theta}}(\widehat{Y}^{\text{test}}, Y^{\text{test}})$, the classification accuracy on the test set according to the chosen metric $C_\theta$ and optimal cut-off $\widehat{\theta}$ determined in Step 5.

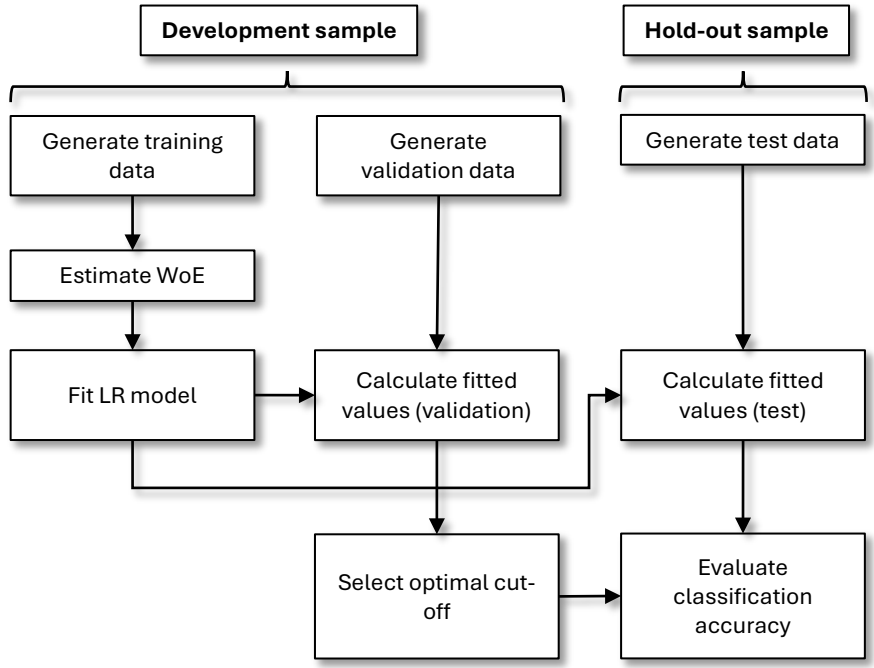

*Figure 1. Outline of the steps followed in each Monte Carlo iteration.*

## 3.2 Comparative configurations

We first consider only four basic configurations with increasing degrees of association between the likelihood of an event occurring and the set of predictors. These configurations have been constructed to investigate the effects of the strength of association and small event rates on both the classification accuracy in terms of the chosen classification metrics ($F_1$ and $P_4$ scores) and the optimal cut-off for classification. The concordance, as measured by the Gini coefficient, is also assessed.

We present in Table 2 the IVs of each predictor and the AIVs for these configurations. Notice that the overall magnitude of the IVs and the AIVs increase as the degree of association increases. For these configurations, we consider four predictors, where the first predictor consists of three bins and the remaining predictors consist of four bins each. This is kept fixed for these configurations to ensure a fair comparison.

*Table 2: Information values (left) of each predictor and the aggregate information value (right) of all predictors for the basic configurations.*

| Configuration | Information value | | | | Aggregate information value |
|---|---|---|---|---|---|
| | $X_1$ | $X_2$ | $X_3$ | $X_4$ | |
| A | 0.0770 | 0.1288 | 0.1595 | 0.0112 | 0.3765 |
| B | 0.6039 | 0.2200 | 0.1992 | 1.2638 | 2.2869 |
| C | 2.2956 | 1.8659 | 0.7290 | 0.4726 | 5.3631 |
| D | 3.9895 | 3.9542 | 3.6617 | 4.9046 | 16.5100 |

We now provide more detailed descriptions of each of these configurations. Several additional configurations are considered later to further investigate the effect of the strength of association in terms of the AIV on classification accuracy. The class-conditional distributions of all four configurations are given in Table 3 and graphically illustrated in Figure 2.

Configuration A represents a situation where there is a weak association between the response and the predictors. Configuration B closely represents a typical real-world situation in credit risk modelling. In terms of the simulation study, this configuration represents a weak to moderate degree of association between the response and the predictors. Configuration C represents a moderate to strong association between the response and the predictors, which may be applicable for certain credit risk models. Lastly, Configuration D corresponds to a situation where there is a very strong association between the response and the predictors. Note that this high degree of association is typically not encountered in practical settings, but it is included in the simulations to gain a better understanding of the relationship between classification accuracy and degree of association.

*Table 3: Class-conditional distributions for the four considered configurations.*

| Configuration | Predictor | $P(X = x_k\|Y = 1)$ | | | | $P(X = x_k\|Y = 0)$ | | | |
|---|---|---|---|---|---|---|---|---|---|
| | | $k=1$ | $k=2$ | $k=3$ | $k=4$ | $k=1$ | $k=2$ | $k=3$ | $k=4$ |
| A | $X_1$ | 0.40 | 0.35 | 0.25 | | 0.30 | 0.33 | 0.37 | |
| | $X_2$ | 0.20 | 0.35 | 0.32 | 0.13 | 0.10 | 0.33 | 0.34 | 0.23 |
| | $X_3$ | 0.10 | 0.50 | 0.25 | 0.15 | 0.15 | 0.60 | 0.20 | 0.05 |
| | $X_4$ | 0.50 | 0.30 | 0.15 | 0.05 | 0.55 | 0.28 | 0.13 | 0.04 |
| B | $X_1$ | 0.38 | 0.51 | 0.11 | | 0.08 | 0.70 | 0.22 | |
| | $X_2$ | 0.18 | 0.34 | 0.29 | 0.19 | 0.05 | 0.33 | 0.32 | 0.30 |
| | $X_3$ | 0.08 | 0.47 | 0.28 | 0.17 | 0.12 | 0.62 | 0.20 | 0.06 |
| | $X_4$ | 0.32 | 0.60 | 0.06 | 0.02 | 0.10 | 0.40 | 0.20 | 0.30 |
| C | $X_1$ | 0.75 | 0.15 | 0.10 | | 0.15 | 0.10 | 0.75 | |
| | $X_2$ | 0.05 | 0.10 | 0.15 | 0.70 | 0.55 | 0.15 | 0.10 | 0.20 |
| | $X_3$ | 0.05 | 0.25 | 0.60 | 0.10 | 0.20 | 0.50 | 0.27 | 0.03 |
| | $X_4$ | 0.09 | 0.10 | 0.15 | 0.66 | 0.30 | 0.20 | 0.10 | 0.40 |
| D | $X_1$ | 0.80 | 0.15 | 0.05 | | 0.05 | 0.20 | 0.75 | |
| | $X_2$ | 0.80 | 0.10 | 0.07 | 0.03 | 0.07 | 0.08 | 0.15 | 0.70 |
| | $X_3$ | 0.10 | 0.05 | 0.15 | 0.70 | 0.80 | 0.10 | 0.07 | 0.03 |
| | $X_4$ | 0.03 | 0.05 | 0.07 | 0.85 | 0.75 | 0.15 | 0.06 | 0.04 |

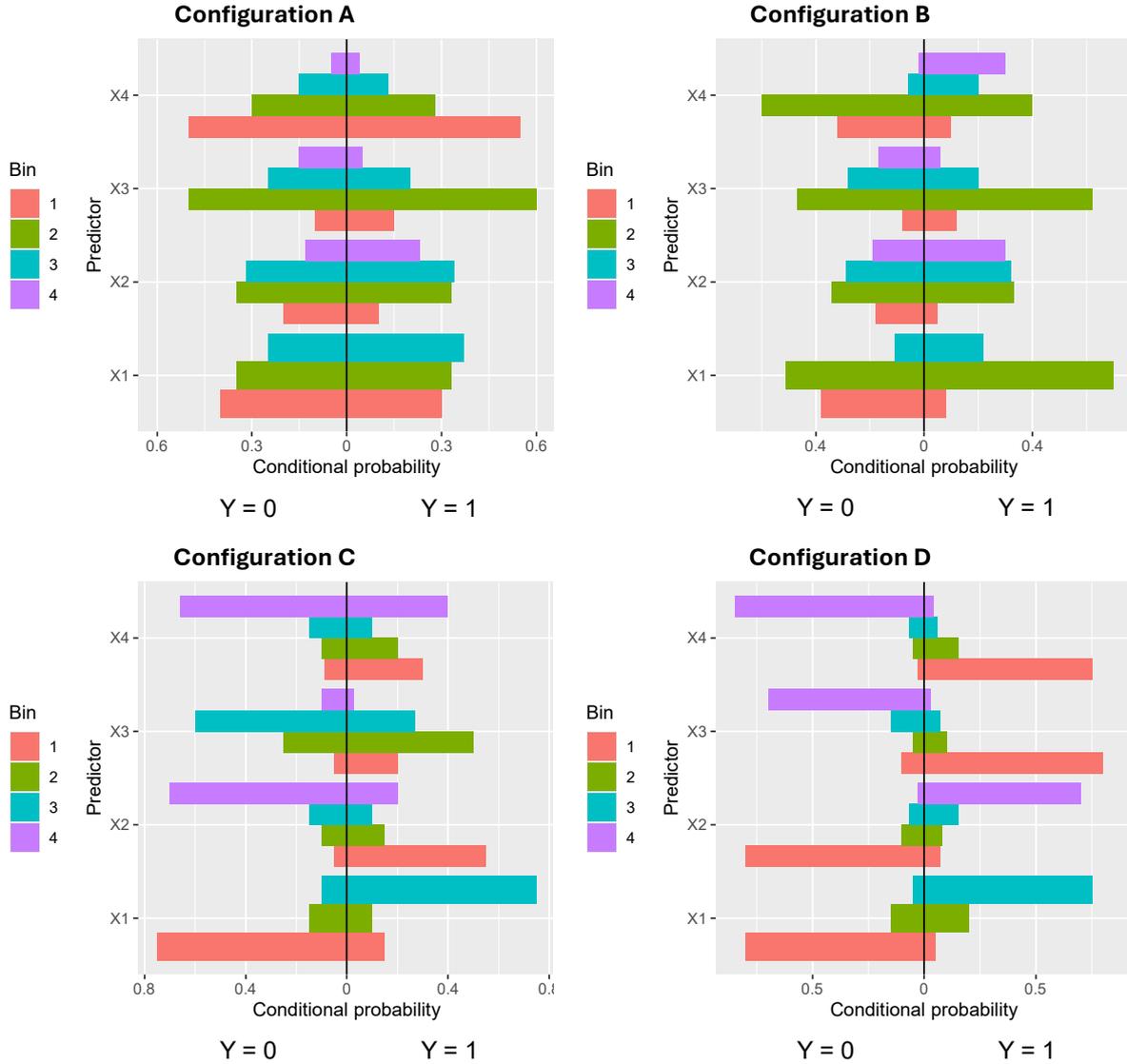

*Figure 2: Visualisation of the class-conditional distributions of the configurations considered.*

### 3.3 Additional configurations

We consider several additional configurations with varying degrees of association between the response variable and predictors. The goal is to gain better insight into the relationship between the strength of association and classification accuracy. Table 4 shows the AIVs of the 16 considered cases, four of which correspond to the basic configurations considered earlier. The configurations are ordered from smallest to largest AIV. Note that $d$, the number of predictors, varies across the configurations.

*Table 4. Aggregate information value (AIV) for each of the 16 cases considered in this section.*

| $d$ | 1 | 4 | 1 | 3 | 1 | 6 | 4 | 3 | 1 | 1 | 4 | 2 | 2 | 2 | 3 | 4 |
|---|---|---|---|---|---|---|---|---|---|---|---|---|---|---|---|---|
| AIV | 0.04 | 0.38 | 0.60 | 0.89 | 1.35 | 2.05 | 2.29 | 3.07 | 3.99 | 4.90 | 5.36 | 5.51 | 7.94 | 8.86 | 11.61 | 16.51 |

These additional configurations will be used to construct general and practical guidelines in terms of logistic regression performance under realistic class imbalance levels in the credit risk environment.

## 4 Simulation results

### 4.1 Classification accuracy

The median (solid line) and quartiles (dashed lines) of the $F_1$ score achieved over the 500 MC repetitions are displayed in Figure 3. The results are displayed separately for the validation and test sets for each of the four basic configurations. Some general observations from these graphs include the following:

- For a given degree of association between the response and the predictors, a maximum median $F_1$ score can be achieved for a specified event rate. The degree of association plays a clear role, where a higher $F_1$ score can be achieved in the case of a stronger association. The level of class imbalance heavily influences the achievable $F_1$ score, where a very small event rate can greatly reduce the classification accuracy. That is, in the case of strong association one can achieve a high $F_1$ score even under severe class imbalance, whereas in the case of very weak association, one can expect a low $F_1$ score even with less severe class imbalance.
- Overall, a significantly larger test set is needed for the median $F_1$ score to stabilise to a similar level achieved on the validation data. Observe that stability is reached on the validation set with fewer than 500 observations in most cases, whereas the median $F_1$ score starts to stabilise at around 1 000 to 1 500 test observations in most cases.
- The variability in the $F_1$ score reduces as the sample size increases. However, the variability in the $F_1$ score increases as the degree of association between the response and the predictors increases. The latter is likely due to the higher variability observed in the optimal classification cut-off determined using the validation set (see Section 4.3).

Considering the realistic configurations (Configurations B and C), great caution should be exercised when the event rate is very low and the sample size is small. For a sample size of less than 500, the classification accuracy for this configuration seems to be inadequate for event rates lower than 5%.

Figure 4 provides analogous results when using the $P_4$ score as the classification metric. While the $P_4$ score is a more comprehensive classification metric than the $F_1$ score, the conclusions remain the same (with only the magnitude of the classification metrics being different).

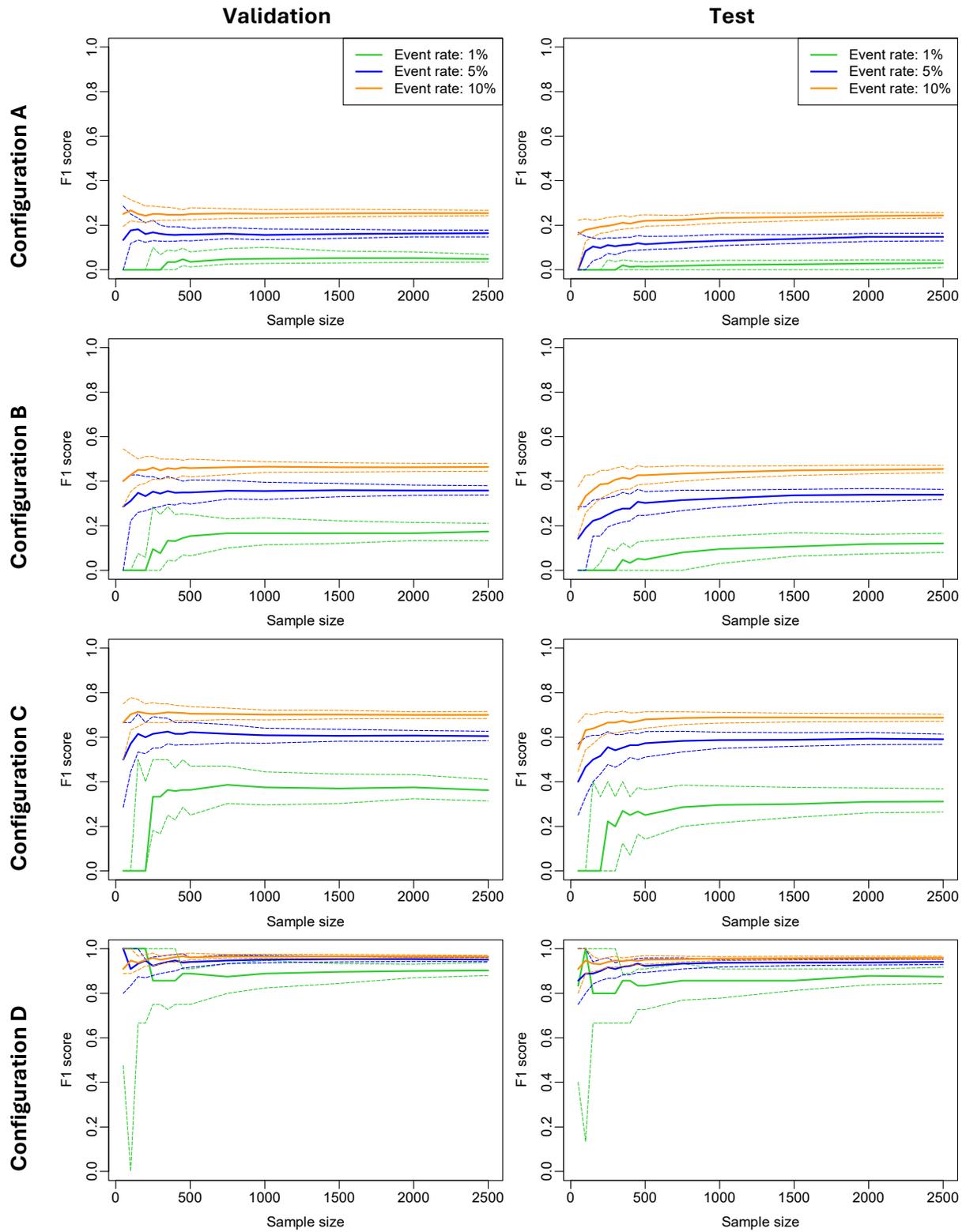

Figure 3: $F_1$ score for the validation and test sets, over the different configurations and for the different fixed event rates. The solid lines represent the median $F_1$ score over the MC repetitions and the associated dashed lines the 25th and 75th percentiles.

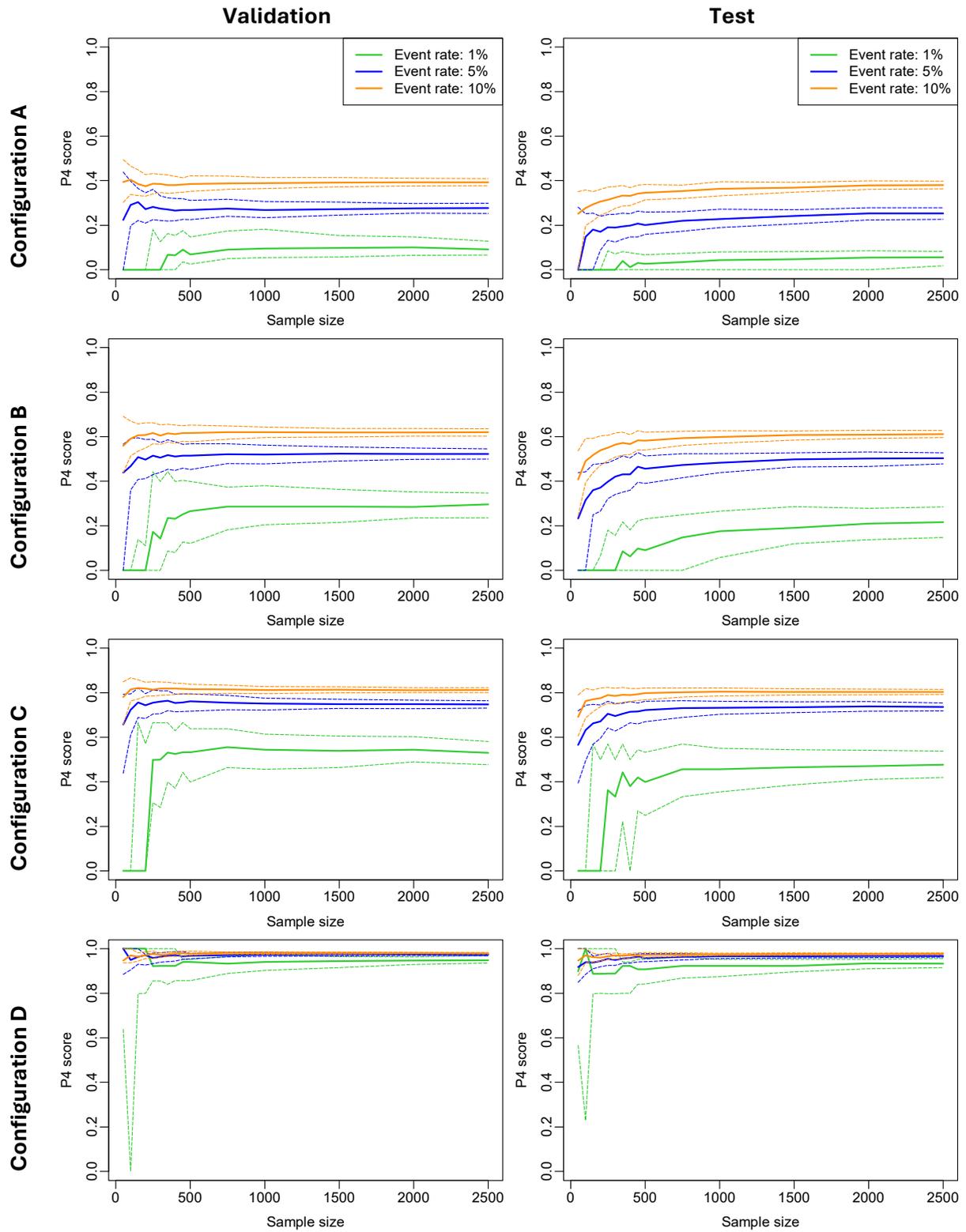

Figure 4: Performance of the $P_4$ score on the validation and testing data sets, over the different configurations and for the different fixed event rates. The solid lines represent the median $P_4$ score over the MC repetitions and the associated dashed lines the 25th and 75th percentiles.

## 4.2 Concordance

We have also studied the effect of a low event rate on concordance as measured by the Gini coefficient. Figure 5 shows the median Gini coefficient (solid line) for each configuration and each event rate, along with the first and third quartiles (dashed lines), obtained from 500 Monte Carlo iterations. We focus on the right-hand panel showing the Gini coefficients obtained on the test sets. The following observations can be made:

- Similar to the $F_1$ score, for a given degree of association, there seems to be a maximum Gini coefficient that can be achieved. Not surprisingly, a higher Gini coefficient can be achieved for higher degrees of association.
- A striking feature visible in Figure 5 is that the Gini coefficient for all considered event rates seems to converge to the same value as the sample size is increased. This is in contrast with the behaviour of the $F_1$ score (see Figure 3) which seems to converge to a different value for each event rate. This suggests that the *ranking* of observations from "good" (most likely to not default) to "bad" (most likely to default) can be done just as effectively under a low event rate as under a high event rate, provided that the sample is large enough. However, as is clear in Figure 3, *classification accuracy* is heavily dependent on the event rate (except if the AIV is unrealistically high as in Configuration D).
- A lower event rate is associated with a lower Gini coefficient, especially for smaller samples. However, related to the previous point, the variability of the Gini coefficient decreases as the sample size is increased, so that the effect of the event rate on concordance diminishes.

## 4.3 Optimal cut-off for classification results

The median (solid line) and quartiles (dashed lines) of the optimal cut-off for classification, determined on the independent validation set over each of the 500 MC repetitions, are shown in Figure 6. The median optimal cut-off, which was determined by optimising the $F_1$ score and $P_4$ score over a grid of values, are compared for the four basic configurations. Some general observations from Figure 6 include the following:

- Although the $P_4$ score is a more comprehensive classification metric than the $F_1$ score, essentially the same optimal cut-offs for classification are chosen by these classification metrics. The behaviour in terms of stabilising and the variability in the optimal cut-offs is also similar for these classification metrics.
- For a given degree of association between the response and the predictors, there seems to be (with some degree of certainty) an optimal cut-off for a specified event rate, which is relatively independent of the sample size, given large enough samples. The strength of association has an influence on the optimal cut-off, where a larger cut-off is selected in the case of stronger association. The level of class imbalance has a significant effect on the optimal cut-off, where a smaller cut-off is selected for a lower event rate.

- The variability in the optimal cut-off increases as the strength of association increases. However, this is expected since a classification model would be able to better distinguish between events and nonevents, thereby allowing for a wider range of cut-offs resulting in similar performance on the performance metrics.
- The behaviour and large variability of the optimal cut-off for Configuration D indicate that the choice of the cut-off might not be very important in the case of such a strong association. This may be due to the predicted values from the logistic regression model being close to 1 for events and close to 0 for nonevents, resulting in many suitable choices for the cut-off.

In the majority of the configurations, the optimal cut-off stabilises between sample sizes of 500 and 1000. This again indicates that caution should be exercised when working with samples with fewer than 500 observations under these levels of class imbalance. Considering that very similar optimal cut-offs are chosen when using the $F_1$ score and $P_4$ score, together with similar conclusions regarding the actual performance measured by these metrics (save for the actual magnitude of the metric), one can consider opting for the simpler and more widely used $F_1$ score in the credit risk modelling environment.

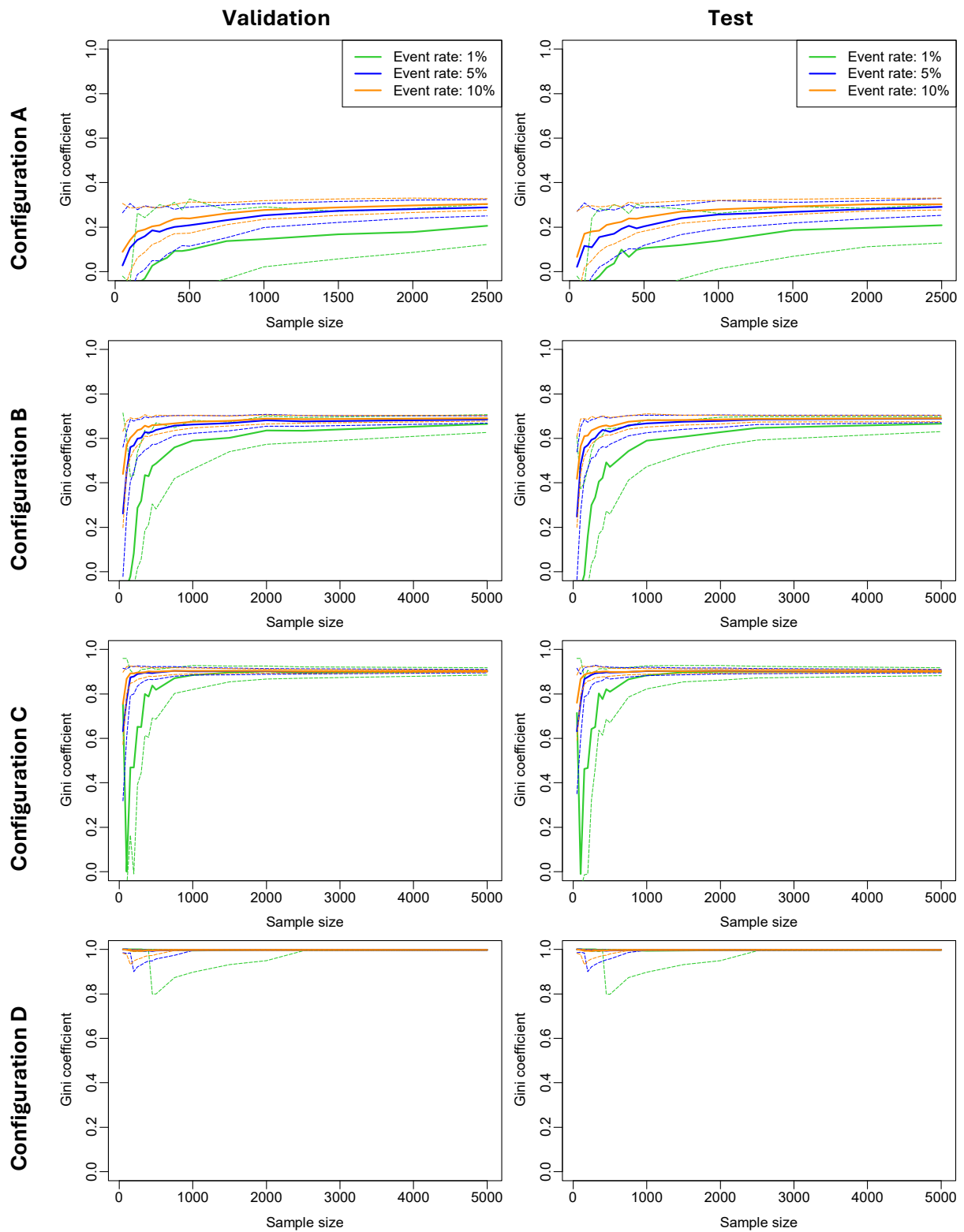

*Figure 5: Gini coefficient for the validation and test sets, over the different configurations and for the different fixed event rates. The solid lines represent the median Gini coefficient over the MC repetitions and the associated dashed lines the 25th and 75th percentiles.*

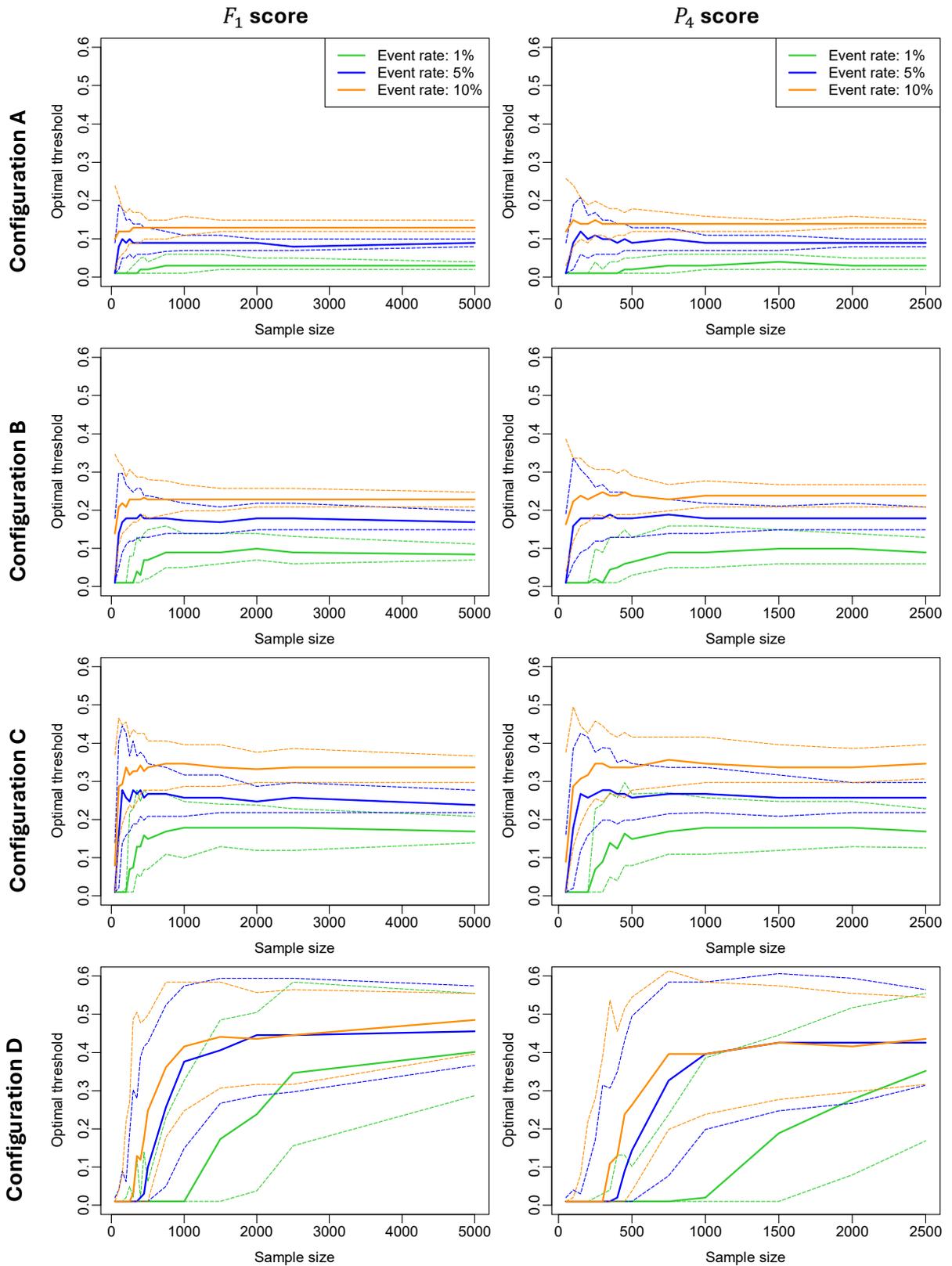

Figure 6: Optimal cut-off for classification using the F1 and P4 scores, over the different configurations and for the different fixed event rates. The solid lines represent the median optimal cut-offs over the MC repetitions and the associated dashed lines the 25th and 75th percentiles.

## 4.4 Additional configurations

Figure 7 shows the median $F_1$ score obtained over 500 Monte Carlo iterations for sample size $n = 250$ for each of the configurations. The dots in the figure correspond to the median $F_1$ score as a function of the AIV. In agreement with the previous findings, the median $F_1$ score increases as the strength of association between the predictors and the response increases. Each solid line shows the least squares fit of a logistic curve.

The vertical bars correspond to 90% intervals from the 5th Monte Carlo percentile to the 95th percentile. The variability of the $F_1$ score for lower event rates tends to be higher than the variability observed in the case of higher event rates. In addition, as the strength of association (in terms of the AIV) increases, the variability of the resulting $F_1$ score seems to increase.

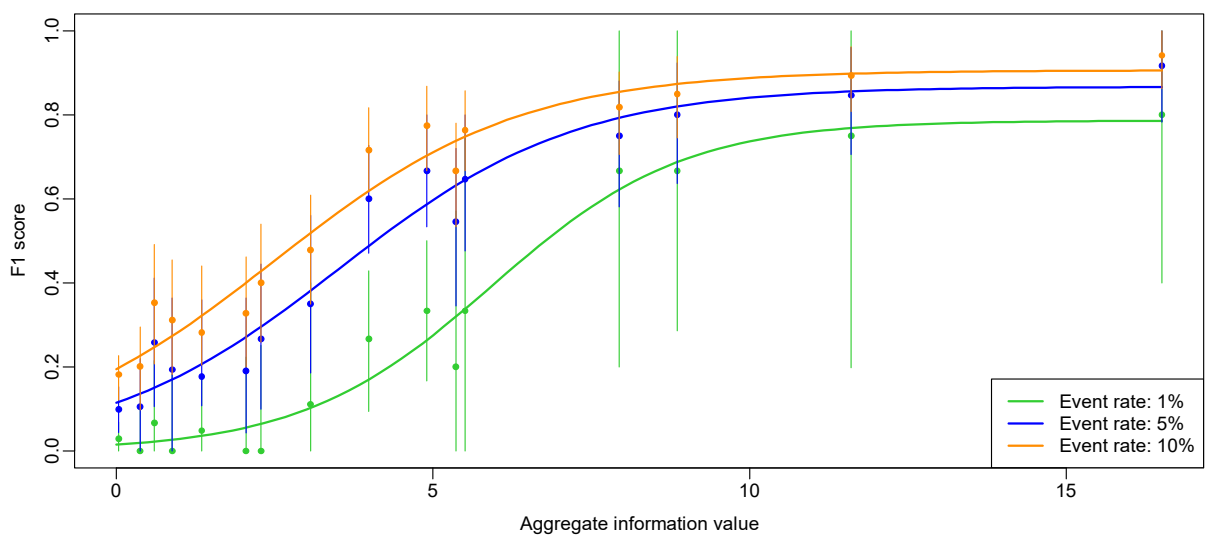

*Figure 7. Median $F_1$ score for each configuration with a fixed event rate and sample size $n = 250$. The vertical lines represent the Monte Carlo variation of each score from the 5th to the 95th percentile.*

Similar conclusions can be made for the cases where the sample sizes are $n = 1\,000$ and $n = 2\,500$ – see Figure 8 and Figure 9. Notice that the variability of the $F_1$ score decreases as the sample size is increased.

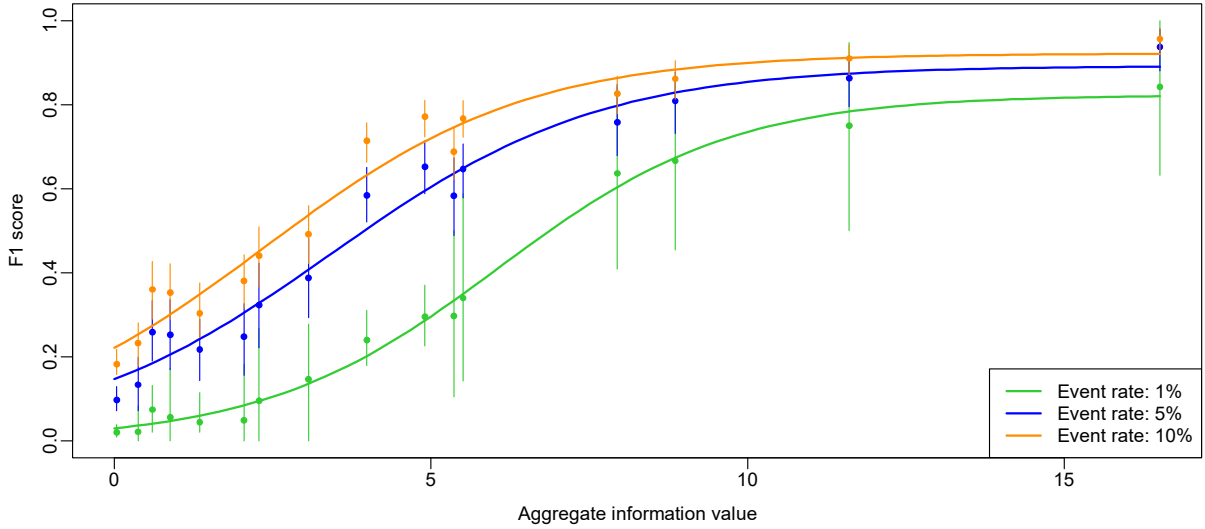

*Figure 8. Median $F_1$ score for all configurations considered with a fixed event rate and sample size $n = 1\,000$. The vertical lines represent the MC variation of the score of each configuration from the 5th to the 95th percentile.*

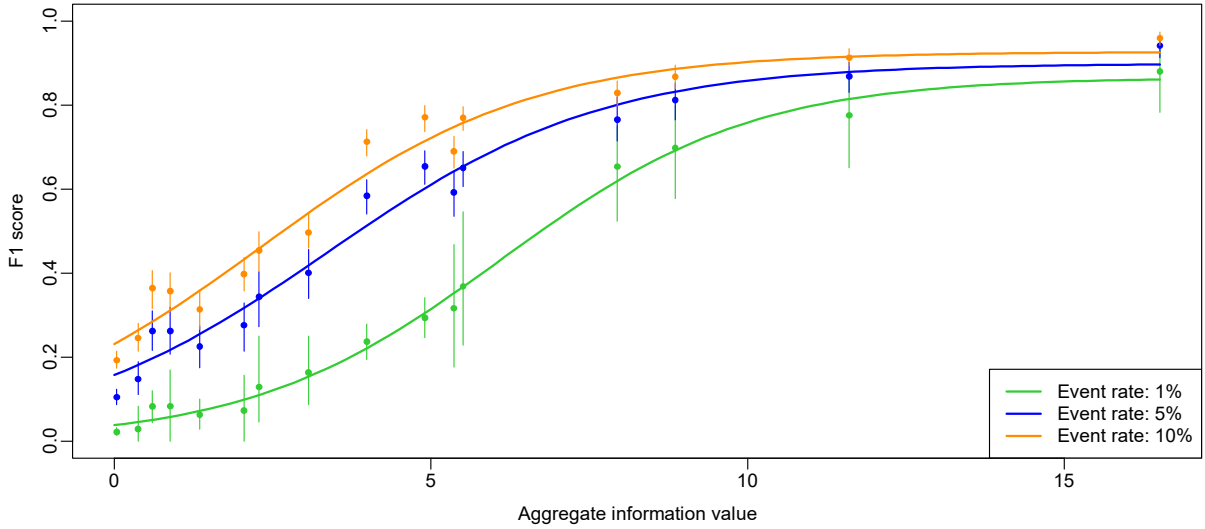

*Figure 9. Median $F_1$ score for all configurations considered with a fixed event rate and sample size $n = 2\,500$. The vertical lines represent the MC variation of the score of each configuration from the 5th to the 95th percentile.*

# 5  Conclusions and practical guidelines

In this study, we have demonstrated by means of a numerical study that a very low event rate can be detrimental to the classification accuracy of a logistic regression classifier. Given the level of class imbalance and AIV, one cannot reasonably expect the classification accuracy to be much higher than a certain maximum value for a specific classification metric. Our simulations indicate that a "safe" sample size, that is where the median classification metric score achieves this maximum, is in the range of 500 to 1 000 observations. The "safe" sample size leans towards the higher end of this range for the lowest event rate of 1%. For the $F_1$ score, these approximate maxima are shown in Table 5.

Table 5. Approximate median $F_1$ score that can be expected for varying levels of the AIV and event rate. These values were approximated using a logistic curve fitted to the data depicted in Figure 9.

| Event rate | Aggregate information value | | | | | | | | | | | | | |
|---|---|---|---|---|---|---|---|---|---|---|---|---|---|---|
| | 0.50 | 1.00 | 1.50 | 2.00 | 2.50 | 3.00 | 3.50 | 4.00 | 4.50 | 5.00 | 5.50 | 6.00 | 6.50 | 7.00 |
| 1% | 0.06 | 0.07 | 0.09 | 0.11 | 0.13 | 0.16 | 0.19 | 0.23 | 0.27 | 0.32 | 0.37 | 0.42 | 0.47 | 0.52 |
| 5% | 0.19 | 0.23 | 0.27 | 0.32 | 0.36 | 0.41 | 0.47 | 0.52 | 0.57 | 0.61 | 0.65 | 0.69 | 0.73 | 0.76 |
| 10% | 0.28 | 0.32 | 0.38 | 0.43 | 0.48 | 0.54 | 0.59 | 0.64 | 0.68 | 0.72 | 0.76 | 0.79 | 0.81 | 0.84 |

Table 5 provides the practitioner with a useful guideline, given that the event rate and the AIV can easily be estimated from sample data. For example, in the situation where the event rate is 1% and the AIV is 2.5, a classification accuracy of 13% (as measured by the $F_1$ score) is reasonable and one cannot expect the logistic regression classifier to do much better, even if the total number of observations (events and nonevents) in the sample is large.

It is also shown that there seems to be a maximum Gini coefficient that can be achieved given a certain AIV. However, unlike the classification accuracy, the Gini coefficient is relatively unaffected by the level of class imbalance. This means that, under a low event rate, observations can be ranked almost as accurately as they can be ranked under a higher event rate. Given that the Gini is the dominant model performance metric used by banks, our results demonstrate that relying only on this metric can be misleading. Considering the typical AIVs and event rates observed in the banking industry, practitioners are recommended to also consider classification accuracy to avoid the risk of underestimating potential credit losses.

Further simulation studies were performed to investigate the performance under a fixed and low number of event cases, although these results are omitted from this paper. An important finding is that the classification accuracy of a logistic regression model cannot be improved by adding only nonevent cases (nondefaults) to a data set. Adding only nonevent cases and keeping the number of event cases fixed lowers the effective event rate and only leads to a deterioration in classification accuracy. Furthermore, adding additional nonevents might bias the data even more given that there could have been a shift in how the drivers relate to the outcome.